\documentstyle[12pt]{article}
\catcode`\@=11
 
\input epsf.tex
\@addtoreset{equation}{section}
\def\theequation{\arabic{equation}}

\def\@normalsize{\@setsize\normalsize{15pt}\xiipt\@xiipt
\abovedisplayskip 14pt plus3pt minus3pt%
\belowdisplayskip \abovedisplayskip
\abovedisplayshortskip  \z@ plus3pt%
\belowdisplayshortskip  7pt plus3.5pt minus0pt}
\def\small{\@setsize\small{13.6pt}\xipt\@xipt
\abovedisplayskip 13pt plus3pt minus3pt%
\belowdisplayskip \abovedisplayskip
\abovedisplayshortskip  \z@ plus3pt%
\belowdisplayshortskip  7pt plus3.5pt minus0pt
\def\@listi{\parsep 4.5pt plus 2pt minus 1pt
            \itemsep \parsep
            \topsep 9pt plus 3pt minus 3pt}}
 
\def\underline#1{\relax\ifmmode\@@underline#1\else
        $\@@underline{\hbox{#1}}$\relax\fi}
\@twosidetrue
\relax
\catcode`@=12
 
\catcode`\@=11

\def\section{\@startsection{section}{1}{\z@}{3.5ex plus 1ex minus
   .2ex}{2.3ex plus .2ex}{\large\bf}}

\def\ps@headings{\def\@oddfoot{}\def\@evenfoot{}
\def\@oddhead{\hbox{}\hfill
        \makebox[.5\textwidth]{\raggedright\ignorespaces --\thepage{}--
        \hfill }}
\def\@evenhead{\@oddhead}
\def\subsectionmark##1{\markboth{##1}{}}
}
 
\ps@headings
 
\catcode`\@=12
 
\relax

\def\figcap{\section*{Figure Captions\markboth
        {FIGURECAPTIONS}{FIGURECAPTIONS}}\list
        {Fig. \arabic{enumi}:\hfill}{\settowidth\labelwidth{Fig. 999:}
        \leftmargin\labelwidth
        \advance\leftmargin\labelsep\usecounter{enumi}}}
 \relax
\def\tablecap{\section*{Table Captions\markboth
        {TABLECAPTIONS}{TABLECAPTIONS}}\list
        {Table \arabic{enumi}:\hfill}{\settowidth\labelwidth{Table 999:}
        \leftmargin\labelwidth
        \advance\leftmargin\labelsep\usecounter{enumi}}}
 \relax
\def\reflist{\section*{References\markboth
        {REFLIST}{REFLIST}}\list
        {[\arabic{enumi}]\hfill}{\settowidth\labelwidth{[999]}
        \leftmargin\labelwidth
        \advance\leftmargin\labelsep\usecounter{enumi}}}
 \relax
 
\catcode`\@=11
 
\def\marginnote#1{}
\newcount\hour
\newcount\minute
\newtoks\amorpm
\hour=\time\divide\hour by60
\minute=\time{\multiply\hour by60 \global\advance\minute by-
\hour}
\edef\standardtime{{\ifnum\hour<12 \global\amorpm={am}%
    \else\global\amorpm={pm}\advance\hour by-12 \fi
    \ifnum\hour=0 \hour=12 \fi
    \number\hour:\ifnum\minute<100\fi\number\minute\the\amorpm}}
\edef\militarytime{\number\hour:\ifnum\minute<100\fi\number\minute}
\def\draftlabel#1{{\@bsphack\if@filesw {\let\thepage\relax
  \xdef\@gtempa{\write\@auxout{\string
    \newlabel{#1}{{\@currentlabel}{\thepage}}}}}\@gtempa
    \if@nobreak \ifvmode\nobreak\fi\fi\fi\@esphack}
     \gdef\@eqnlabel{#1}}
\def\@eqnlabel{}
\def\@vacuum{}
\def\draftmarginnote#1{\marginpar{\raggedright\scriptsize\tt#1}}
\def\draft{\oddsidemargin -.5truein
        \def\@oddfoot{\sl preliminary draft \hfil
        \rm\thepage\hfil\sl\today\quad\militarytime}
        \let\@evenfoot\@oddfoot \overfullrule 3pt
        \let\label=\draftlabel
        \let\marginnote=\draftmarginnote
   
\def\@eqnnum{(\theequation)\rlap{\kern\marginparsep\tt\@eqnlabel}%
\global\let\@eqnlabel\@vacuum}  }
\def\preprint{\twocolumn\sloppy\flushbottom\parindent 1em
        \leftmargini 2em\leftmarginv .5em\leftmarginvi .5em
        \oddsidemargin -.5in    \evensidemargin -.5in
        \columnsep 15mm \footheight 0pt
        \textwidth 250mmin      \topmargin  -.4in
        \headheight 12pt \topskip .4in
        \textheight 175mm
        \footskip 0pt
\def\@oddhead{\thepage\hfil\addtocounter{page}{1}\thepage}
        \let\@evenhead\@oddhead \def\@oddfoot{} \def\@evenfoot{} 
}
\def\titlepage{\@restonecolfalse\if@twocolumn\@restonecoltrue\onecolumn
     \else \newpage \fi \thispagestyle{empty}\c@page\z@
        \def\thefootnote{\fnsymbol{footnote}} }
\def\endtitlepage{\if@restonecol\twocolumn \else  \fi
        \def\thefootnote{\arabic{footnote}}
        \setcounter{footnote}{0}}  %\c@footnote\z@ }
\catcode`@=12
\relax
\def\ps@headings{\def\@oddfoot{}\def\@evenfoot{}
\def\@oddhead{\hbox{}\hfill
        \makebox[.5\textwidth]{\raggedright\ignorespaces --\thepage{}--
        \hfill }}
\def\@evenhead{\@oddhead}
\def\subsectionmark##1{\markboth{##1}{}}
}
 
\ps@headings
 
\relax

\def\firstpage#1#2#3#4#5#6{
\begin{document}
%\draft
%%%%%%%%%%%%%%%%% MACROS %%%%%%%%%%%%%%%%%%
\def\beq{\begin{equation}} 
\def\eeq{\end{equation}} 
\def\bea{\begin{eqnarray}} 
\def\eea{\end{eqnarray}} 
\def\bq{\begin{quote}} 
\def\eq{\end{quote}}
\def\ra{\rightarrow} 
\def\lra{\leftrightarrow} 
\def\ups{\upsilon}
\def\bq{\begin{quote}} 
\def\eq{\end{quote}}
\def\ra{\rightarrow} 
\def\un{\underline}
\def\ov{\overline}
\newcommand{\cm}{Commun.\ Math.\ Phys.~}
\newcommand{\ibar}{\bar{\imath}}
\newcommand{\jbar}{\bar{\jmath}}
\newcommand{\F}{{\cal F}}
\renewcommand{\L}{{\cal L}}
\newcommand{\A}{{\cal A}}
\newcommand{\eps}{\epsilon}
\newcommand{\th}{\theta}
\newcommand{\notl}{l\hspace{-.20cm}-}
\newcommand{\notj}{j\hspace{-.20cm}_-}
\newcommand{\PR}{{\it Phys. Rev. }}
\newcommand{\PL}{{\it Phys. Lett. }}
\newcommand{\NP}{{\it Nucl. Phys. }}
\newcommand{\PRL}{{\it Phys.\ Rev.\ Lett.}}
\def\154{\frac{15}{4}}
\def\153{\frac{15}{3}}
\def\32{\frac{3}{2}}
\def\254{\frac{25}{4}}
\begin{titlepage}
\nopagebreak
\title{\begin{flushright}
        \vspace*{-1.8in}
        {\normalsize NTUA-65/97}\\[-9mm]
        {\normalsize IOA-TH/97-005}\\[3mm]
        \vspace*{-1.2cm}
        {\normalsize CERN-TH/97-268}\\[-9mm]
\end{flushright}
\vfill
{#3}}
\author{\large #4 \\[1.0cm] #5}
\maketitle
\vskip -7mm     
\nopagebreak 
\begin{abstract}
{\noindent #6}
\end{abstract}
\vfill
\begin{flushleft}
\rule{16.1cm}{0.2mm}\\[-3mm]
$^{\star}${\small Research supported in part by 
TMR contract ERBFMRX-CT96-0090, $\Pi$ENE$\Delta$-1170/95 and $\Pi$ENE$\Delta$-15815/95}\\
%e-mails: George.Leontaris@cern.ch, ntrac@central.ntua.gr\\ 
September 1997
\end{flushleft}
\thispagestyle{empty}
\end{titlepage}}
 
\def\simlt{\stackrel{<}{{}_\sim}}
\def\simgt{\stackrel{>}{{}_\sim}}
\date{}
\firstpage{3118}{IC/95/34}
{\large\bf  
Modular Weights, $U(1)$'s and Mass Matrices$^{\star}$} 
{G.K. Leontaris $^{\,a,b}$ and N.D. Tracas$^{\,c,b}$}%\\[-3mm]
{\normalsize\sl
$^a$Theoretical Physics Division, Ioannina University,
{}GR-451 10 Ioannina, Greece\\[-3mm]
\normalsize\sl
$^b$ CERN, Theory Division, Geneva\\[-3mm]
\normalsize\sl
$^c$ Physics Department, National Technical University,
157 80 Athens, Greece.}
{We derive the scalar mass matrices in effective supergravity models 
with the standard gauge group augmented
 by a $U(1)_F$ family symmetry. Simple relations between $U(1)_F$ charges and
modular weights of the superfields are derived and used to express
the matrices with a minimum number of parameters. The model predicts 
a branching ratio for the $\mu\rightarrow e\gamma$ process close
to the present experimental limits.}
{.}

\newpage
The Minimal Supersymmetric Standard Model (MSSM) emerges as the most
natural extension of the Standard Model (SM) in the context of the
unification of all interactions. Although supersymmetric models
solve  the hierarchy problem, the plethora of arbitrary
parameters requires a further step beyond the MSSM. The $N=1$ 
supergravity coupled to matter stands promising
%%%%%%%%%%%%%%%%%%%
\cite{SUGRA}.
%%%%%%%%%%%%%%%%%%%
Yet, there are  many essential parameters  (Yukawa couplings, content of the chiral
multiplets, etc.) to be chosen by the model builder. In this scene,  string theory
appears the only known candidate theory that can in principle predict all the
required parameters.  String theory puts rather strong constraints on many of the
parameters of the resulting $N=1$ effective supergravity, which appears as its 
low-energy limit. Thus, the kinetic terms must have a certain structure, the Lagrangian
should obey the string duality symmetries, while  several constraints are imposed
on the superpotential and the Yukawa couplings
%%%%%%%%%%%%%%%%%%%%
\cite{duality}.
%%%%%%%%%%%%%%%%%%%%
 
The subject of this letter is to reproduce the observed family hierarchy of the
fermion masses and moreover to predict the corresponding mass matrices 
in the scalar supersymmetric sector. This is done in the context of residual stringy
$U(1)$ symmetries
%%%%%%%%%%%%%%%%%%%%
\cite{u1} 
%%%%%%%%%%%%%%%%%%%%
left from the large gauge group at a high scale. In particular,
combining modular invariance constraints and $U(1)$ invariance of the superpotential,
%and the scalar sector obtained from the K\"ahler potential,
the scalar mass matrices 
are given in terms of powers of an expansion parameter $\langle\theta\rangle/M$, 
where $\langle\theta\rangle$ is  the vacuum expectation value of a singlet field and $M$ is a
high (string) scale. These powers are written in terms of modular weight differences.
Further, the consequences in the lepton flavour non-conserving reaction
$\mu\ra e\gamma$ are  examined.  Its branching ratio is found close to the present
experimental limits.

We start with a quick review of the $N=1$ supergravity, which introduces 
a real gauge-invariant K\"ahler function with the general form
%%%%%%%%%%%%%%%%%%%%
\cite{kahler}
%%%%%%%%%%%%%%%%%%%
\beq
G(z,\bar z) = {\cal K}(z,\bar z) + \log{\mid {\cal W}(z)\mid}^2
\label{kahlerfun}
\eeq
where ${\cal K}(z,\bar z)$ is the K\"ahler potential and ${\cal W}(z)$ is the
superpotential. Denoting $(\Phi,Q)$ by $z$, where $\Phi$
stands for the dilaton field $S$ and other moduli 
$T_i$, while $Q$ represents the chiral superfields, 
the K\"ahler potential at tree level can be written as follows
%%%%%%%%%%%%%%%%%%
%\cite{kahlerexp}
%%%%%%%%%%%%%%%%%
\begin{eqnarray}
K(\Phi,\bar{\Phi},Q,\bar Q)&=&-\log(S+\bar S)-\Sigma_nh_n\log(T_{n}+\bar T_{n})+
Z_{\bar ij} (\Phi,\bar\Phi)\bar Q_{\bar i}e^{2V}Q_j+\cdots
\label{kahlerpot}
\end{eqnarray}
The superpotential ${\cal W}(z)$ is a holomorphic
function of the chiral superfields $Q_i$ and at the tree level is given by
\beq
{\cal W}({\Phi},Q) = \frac 13\lambda_{ijk}({\Phi})Q_iQ_jQ_k +
 \frac 12 \mu_{ij}(\Phi)Q_iQ_j + \cdots  \label{sp}
\eeq
In both (\ref{kahlerpot}) and (\ref{sp}), $\cdots$ stand for 
possible non-renormalizable contributions. Terms bilinear in the fields $Q_i$  
refer in fact to an effective Higgs  mixing term.

Now under the modular symmetries, the moduli transform as
$T\rightarrow (aT -\imath b)/(\imath cT+d)$, 
where $a,b,c,d$ constitute the entries of the $SL(2,{\bf Z})$ group elements 
with $a,b,c,d\in {\bf Z}$ and $ad-bc=1$. These imply the 
following transformation rules
%%%%%%%%%%%%%%%
\cite{trans}
%%%%%%%%%%%%%%%
\begin{eqnarray}
Q_i \rightarrow  Q_i\prod_k t_k^{n_i^k},\quad
Z_{i\bar j} \rightarrow  Z_{i\bar j}\prod_k \left(t_k\right)^{-n^k_i}
                                             \left({\bar t}_k\right)^{-n^k_j},
\quad
{\cal W} \rightarrow  \prod_k t_k^{n^k_W}
{\cal W},
\label{trans}
\end{eqnarray}
where we have introduced the notation $t_k=\imath c_kT_k+d_k$. The exponent
$n^k_i$ is the modular weight of $Q_i$ with respect to the modulus $T_k$.

Let us now introduce into the K\"ahler function non-renormalizable
terms through two fields, $\theta$ and ${\bar\theta}$, which are 
singlets under the low energy standard gauge group, while they carry 
charges $q_{\theta}=- q_{\bar\theta}$ under the $U(1)_F$ family group. The lower order (in $Q_i$'s)
non-renormalizable terms can be written in the form
\beq
K_{r_{i\bar j}}^{i\bar j} \left(\frac{\langle\theta\rangle}{M_1}\right)^{r_{i\bar j}}
Q_i\bar Q_{\bar j}+
K_{\tilde r_{i\bar j}}^{i\bar j}
\left(\frac{\langle\bar\theta\rangle}{M_2}\right)^{\tilde r_{i\bar j}}
                                           Q_i\bar Q_{\bar j}.
\label{n-r-terms}
\eeq
These terms should be invariant under the $U(1)_F$ symmetry.
Assigning $U(1)_F$ charges $q_i$ for the matter fields one gets
\beq
q_i+q_j+q_{{\theta}}r_{ij}=0,\quad q_i+q_j+q_{{\bar\theta}}\tilde r_{ij}=0.
\eeq
Similar non-renormalizable terms could also appear in the superpotential. 

After this short review we come to the 
mass matrix textures. The $SU(2)_L$ invariance, together with the requirement
to have symmetric mass matrices, leads us to assign the same $U(1)_F$ charge
to all quark members of the same family $q_i$, while the same should be
applied to the leptons of the same family $l_i$.
The full anomaly-free Abelian group involves an additional family-independent 
component  and with this freedom we may
make $U(1)_F$ traceless without any loss of generality. Thus
$q_1+q_2+q_3=0$ and $l_1+l_2+l_3=0$.

If the light Higgs $H_{2}$, responsible for the masses of the up
quarks, and $H_{1}$, responsible for the down quarks and leptons
have $U(1)_F$ charge, so that only
the (3,3) renormalizable Yukawa couplings to $H_{2}$ and $H_{1}$ are
allowed, namely
\begin{equation}
2q_3+h_2=0,\quad {\mbox {and}}\quad 2l_3+h_1=0,
\label{traceless}
\end{equation}
only the (3,3) element of the associated mass matrix
will be non-zero. The remaining entries are generated    
when the $U(1)_F$ symmetry is broken. A straightforward consequence
of this fact is the equality of the two Higgs $U(1)_F$ charges
($h_1=h_2$), since $H_1$ provides also the mass to the bottom quark
while we have assumed equal $U(1)_F$ charges within a family.

A general non-renormalizable relevant term in the superpotential is of the form
\beq
Y_{ij}Q_iu^c_jH_2 \left({\frac{\theta}{M}}\right)^{x_{ij}}.\label{nrup}
\eeq
Owing to $U(1)_F$ invariance of the superpotential, we have the constraint
\beq
q_i +q_j +h_2 +x_{ij}q_\theta = 0
\label{nrU1}
\eeq
and similarly for the parameter $\bar\theta$ terms.
%where $q_{\th}$ is the charge of the singlet field $\th$ and the
%$\pm$ sign depends on whether $\th$ or $\bar\th$ appears in the term.
The allowed powers of non-renormalizable terms in each       
entry are determined by the charges ($q_{\theta}=-q_{\bar\theta}$)
\begin{eqnarray}
x_{ij}&=&
|\frac{q_2-q_3}{q_\th}|
\left(\begin{array}{ccc}
2|a+1| & |a| & |a+1| \\
|a| & 2 & 1 \\
|a+1| & 1  &0 
\end{array}
\right)
\label{eq:p11}
\end{eqnarray}
where $a=3q_3/(q_2-q_3)$, %$|p_{\pm}|=|q_2-q_3|$ 
and we have used the condition (\ref{traceless}).  
Suppressing unknown Yukawa couplings $Y_{ij}$ and their phases, which
are all expected to be of order 1, we arrive at the following
mass matrices
\begin{eqnarray}
{m_U}&\approx& \left(
\begin{array}{ccc}
\epsilon^{2|a+1|} &
     \epsilon^{|a|} &
           \epsilon^{|a+1|}
\\
\epsilon^{|a|} &
     \epsilon^{ 2 } &
          \epsilon 
\\
\epsilon^{|a+1|} &
      \epsilon & 1
\end{array}
\right){m_t}\\
\label{eq:mu0}
{m_D}&\approx& \left (
\begin{array}{ccc}\tilde{\epsilon}^{2|a+1|} &
      \tilde{\epsilon}^{|a|} &
           \tilde{\epsilon}^{|a+1|} 
\\
\tilde{\epsilon}^{|a|} &
         \tilde{\epsilon}^{ 2 } &
              \tilde{\epsilon} 
\\
\tilde{\epsilon}^{|a+1|} &
\tilde{\epsilon} &1
\end{array}
\right){m_b}
\label{eq:md0}
\end{eqnarray} 
where $\tilde{\epsilon} = (\frac{\langle\theta \rangle}{M_1})^{|(q_2-q_3)/q_{\th}|}$,
$\epsilon=(\frac{\langle\theta \rangle}{M_2})^{|(q_2-q_3)/q_\th|}$ 
($M_1$ and $M_2$ being two high scales).
The charged lepton mass matrix may similarly be determined.
The equality $h_1=h_2$, together with (\ref{traceless}) has also 
the consequence $q_3=l_3$, which implies
the successful relation $m_b=m_{\tau}$ at unification. We then get
\begin{equation}
{m_L}\approx \left (
\begin{array}{ccc}
\tilde{\epsilon}^{2|a +b|} &
        \tilde{\epsilon}^{|a|} &
              \tilde{\epsilon}^{|a +b|} 
\\
\tilde{\epsilon}^{|a|} &
        \tilde{\epsilon}^{2|b|} &
             \tilde{\epsilon}^{|b|} 
\\
\tilde{\epsilon}^{|a +b|} &
       \tilde{\epsilon}^{|b|} &1
\end{array}
\right){m_{\tau}}
\label{ml0}
\end{equation}
where $b =(l_2-q_3)/(q_2-q_3)$.

%%%%%%%%%%%%%%%%
The powers of the above matrices can be written in terms of the
modular weights as follows.
As we have already discussed in the introduction, the superpotential transforms
covariantly under the modular symmetry. Let us denote
by $n_{Q_i},n_{u_i},n_{d_i},n_{h_2},n_\th$ the modular weights for the  
corresponding fields  with respect to a certain modulus. 
For the  non-renormalizable term of the form (\ref{nrup}),
the modular weights obey the equation
$n_{Q_i}+n_{u_j}+n_{h_2}+x_{ij}n_{\theta}=n_{{\cal W}}$. 
Combining this
relation with the $U(1)_F$ invariance and the fact that
$\sum_{i=1}^3q_i=0$, we obtain the general formula
\beq
q_j = \frac{q_{\th}}{3n_{\th}}\sum_{i}n_{Q_{ji}}=
      \frac{q_{\th}}{3n_{\th}}\sum_{i}n_{u_{ji}}=
      \frac{q_{\th}}{3n_{\th}}\sum_{i}n_{d_{ji}}
\label{q-n}
\eeq
where $n_{Q_{ji}}=n_{Q_j}-n_{Q_i}$ and correspondingly for $n_{u_{ji}}$ and
$n_{d_{ji}}$. The third equality comes from the down-quark mass matrix non
renormalizable contributions corresponding
to a term like (\ref{nrup}).
Similar relations hold for the lepton modular weights.   
Using the above relation, we may obtain an elegant form of the
matrix (\ref{eq:p11}), which expresses the powers of the
allowed non-renormalizable entries only in terms of modular
weight differences
\cite{sav}.
 We obtain
\begin{eqnarray}
x_{ij}=
\frac{1}{n_{\th}}
\left(\begin{array}{ccc}
2n_{Q_{31}} &n_{Q_{31}}+n_{Q_{32}} &n_{Q_{31}} \\
n_{Q_{31}}+n_{Q_{32}} & 2n_{Q_{32}} &  n_{Q_{32}} \\
n_{Q{31}} & n_{Q{32}}  &0
\end{array}
\right)
\label{eq:p2}
\end{eqnarray}
The positivity of the entries requires the conditions
$n_{Q_{31}}n_\th>0 $ and $n_{Q_{32}}n_\th>0$.
We can also express the powers of the matrix (\ref{eq:mu0}) in terms of 
modular weight difference. This is easily done by expressing the 
parameter $a$ in the form
\beq
a = \frac{n_{Q_{13}}+n_{Q_{23}}}{n_{Q_{23}}}.
\eeq
{}From (\ref{eq:p2}) we conclude that the hierarchical 
fermion mass spectrum  requires all three $n_{Q_i}$'s
to be different.
Models with equal $n_{Q_i}$'s, but different $q_i$'s (necessary
to create hierarchy), require $q_\th=0$.
In this case the $U(1)_F$
charges are not related to the modular weights and the
constraint (\ref{q-n}) does not hold.

We next turn to the lepton fermion mass matrix. 
The phenomenological constraint $l_3=q_3$ imposes the following
relation on the modular weights of the quark and lepton
generations
\beq
n_{L_{13}} - n_{Q_{13}} = n_{Q_{23}} - n_{L_{23}} \equiv \delta.
\eeq
As a result, the $U(1)_F$ structure permits to express the powers $y_{ij}$ of
the lepton term
\beq
L_ie_jH_1
\left(\frac{\stackrel{(-)}{\theta}}{M}\right)^{y_{ij}}
\label{nrl}
\eeq
by  the following matrix
\begin{eqnarray}
y_{ij}=
\frac{1}{n_{\th}}
\left(\begin{array}{ccc}
2(n_{Q{13}}+\delta) & n_{Q_{13}}+n_{Q_{23}} & n_{Q_{13}}+\delta \\
n_{Q_{13}}+n_{Q_{23}} & 2(n_{Q_{23}}-\delta) &  n_{Q_{23}}-\delta \\
n_{Q_{13}}+\delta & n_{Q_{23}} -\delta &0
\end{array}
\right)
\label{eq:p3}
\end{eqnarray}
whilst the corresponding constraints for the positivity of the entries are
$n_{L_{13}}n_\th>0$ and $n_{L_{23}}n_\th>0$.
The powers of the matrix (\ref{ml0}) can also be expressed in the same way
by writing $b$ in the form
\beq
b=\frac{n_{Q_{23}}-\delta}{n_{Q_{23}}}.
\eeq
 
We now turn to the scalar part. At the tree level the scalar mass
matrices receive contributions only along the diagonal, since
terms of the form $\propto Q_iQ_i^*$ have zero $U(1)_F$ charge.
Using powers of the fields $\th,\bar\th$ scaled by
the $M$, we may fill in the remaining entries.
It can easily be seen that the allowed $U(1)_F$ structure of the powers
in the scalar mass term is
\begin{eqnarray}
\frac{1}{n_{\th}}
\left(\begin{array}{ccc}
0 & n_{Q_{12}} & n_{Q_{13}} \\
n_{Q_{12}} &0 &  n_{Q_{23}} \\
n_{Q_{13}} & n_{Q_{23}}  &0
\end{array}
\right)
\label{eq:3}
\end{eqnarray}
Thus, the powers of the parameters $\langle\th\rangle,\langle\bar\th\rangle$
are simply determined by the differences $n_{Q_{ij}}$ for
the squark matrix and similarly for the 
sleptons (remember that since the
$U(1)_F$ charge is the same within a family, (\ref{q-n}) tells us that $n_{Q_{ij}}=
n_{u_{ij}}=n_{d_{ij}}$).

Using again the parameters $a$ and $b$ entered in the fermion mass
matrices, we can express the squark mass matrix in the form
\beq
m_{\tilde Q}^2 \approx
\left(
\begin{array}{ccc}
{1} & 
\eps^{\mid a + 6\mid}
&\eps^{\mid a + 1\mid } \\
\eps^{\mid a  + 6\mid} 
& 1 &
\eps\\
\eps^{\mid a + 1\mid } & \eps & 1
\end{array}
\right)m_{3/2}^2
\eeq
where $m_{3/2}$ is the gravitino mass. Similarly, for the sleptons we obtain
\beq
m^2_{\tilde{l}_{L,R}} \approx
\left (
\begin{array}{ccc}
{1} & \tilde\eps^{\mid a + 6 b\mid  }
&\tilde\eps^{\mid a + b\mid} \\
\tilde\eps^{\mid a + 6 b \mid } & {1} &
\tilde\eps^{\mid b \mid}\\
\tilde\eps^{\mid a + b\mid } & \tilde\eps^{\mid b\mid} & 1
\end{array}
\right)m_{3/2}^2
\eeq
Obviously in the case of $b =1$ the two matrices are identical
since this case corresponds to equal $U(1)_F$ charges in
the quark and the leptonic sector,  $l_i = q_i$. In fact, it can be checked that
the phenomenological analysis of the fermion mass spectrum 
allows two values of $b$, namely  $b=1$ or $1/2$
%%%%%%%%%%%%%%%%%%%%%%%
\cite{IR},
%%%%%%%%%%%%%%%%%%%%%%%
while $\epsilon \approx 0.053$ and $\tilde\epsilon \approx 0.23$.

The above results show that $U(1)_F$ symmetries necessarily lead to low energy 
models where the Yukawa and its  corresponding scalar mass 
matrices are not simultaneously diagonalized. As a result, flavour violation 
is possible and in general one should check whether such models can pass also the
flavour violation tests. One of the most popular flavour non-conserving processes is
the $\mu \ra e \gamma$ decay. We have calculated the branching 
ratio for this process in order to compare it with the present experimental limits.
This calculation requires the diagonalization of the $6\times 6$
scalar mass matrix
\beq
\tilde{M}_{\tilde l}^2=
\left(\begin{array}{cc}
m_{{\tilde l}_L}^2 & A_{l}\langle H_1\rangle +m_L\mu \tan\beta  \\
(A_{l}\langle H_1\rangle +m_L\mu \tan\beta)^\dagger  &m_{{\tilde l}_R}^2
\end{array}
\right)
\label{eq:31}
\eeq
Here, as usual, $A_l$ is the trilinear parameter entering the scalar potential,
$\mu$ is the Higgs mixing term and  $\tan\beta$ is the Higgs vev ratio.
Since lepton mass matrices are symmetric, left and right
diagonalizing matrices coincide. Further, due to the properties of the $U(1)_F$
symmetry of the model, left ($m_{\tilde{l}_L}^2$) and right ($ m_{\tilde{l}_R}^2$)
scalar mass matrices are the same. 
Moreover, here we restrict on the case of small $\tan\beta$ regime, where
the chirality changing diagrams are suppressed. 
In the general case, of course, and in the large $\tan\beta$ scenario, they become
important. We have considered contributions from one loop graphs involving 
neutralino-charged slepton or chargino-sneutrino states in the loop. 
The diagrams of this process are shown in Fig. 1.
We have diagonalized the lepton and the slepton mass matrices and found the
corresponding amplitude for each neutralino/chargino graph. Then by diagonalizing
the Wino mass matrices we evaluated the total amplitude and the
branching ratio. For sensible values of $m_{3/2}\sim m_{1/2}\sim {\cal O}(m_W)$
(initial
values for the scalar and gaugino masses respectively) and standard GUT initial
conditions for gauge couplings, the value of the BR$_{\mu\rightarrow e\gamma}$ can
reach the order of $10^{-12}$.
Thus, this rare decay gives the opportunity  to
test the viability of the above $U(1)_F$-like  model in
future experiments.

In conclusion, we have considered the scalar mass matrices in supergravity
models with the standard $SU(3)\times SU(2)_L\times U(1)_Y$ gauge group 
augmented by a $U(1)_F$ family symmetry. Using modular invariance of the
K\"ahler potential and the superpotential, we have derived certain relations
between $U(1)_F$ charges and the modular weights of the fields. As a result,
the scalar mass matrix entries are found to depend only on certain powers, which are
proportional to the difference of modular weights.
We have calculated, as an example, the process $\mu\rightarrow e\gamma$, which is found,
for a wide range of the parameter space ($\tan\beta$, $m_{3/2}$, $m_{1/2}$),
to be very close to the present experimental limits. This fact makes it
possible to test such theories in near future-experiments.
\begin{figure}
\begin{center}
\leavevmode
\epsfysize=3in \epsfbox{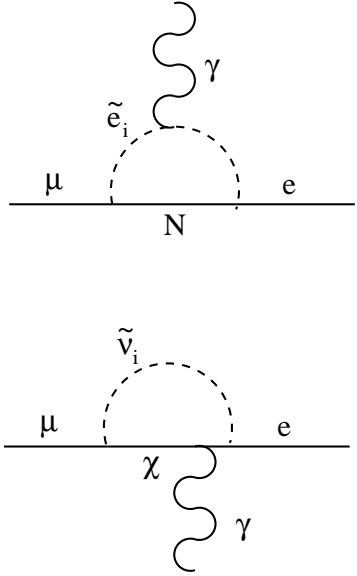}
\caption[]{
 The $\mu\ra e \gamma$ decay via supersymmetric graphs. }
\end{center}
\end{figure}


\newpage 


\begin{thebibliography}{99}
\bibitem{SUGRA}
E. Cremmer, S. Ferrara, L. Girardello and A. Van Proyen, 
\NP {\bf B212}(1983)413;\\
J. Bagger, \NP{\bf B211}(1983)302.
%%%%%%%%%
\bibitem{duality}
S. Ferrara, D. L\"ust and S. Theisen, \PL{\bf B233}(1989)147;\\
S. Ferrara, C. Kounnas, and F. Zwirner,\NP {\bf 365}(1991)431;\\
I. Antoniadis, K.S. Narain and T. Taylor, \PL{\bf B267}(1991)37;\\
S. Kalara, J. Lopez and D.V. Nanopoulos, \PL{\bf B269}(1991)84;\\
L. Iba\~n\'ez and D. L\"ust, \NP{\bf B382}(1992)305;\\
P. Bin\'etruy and E. Dudas, \NP{\bf B442}(1995)21.
%%%%%%%%%%%
\bibitem{u1} 
C. D. Froggatt and H. B. Nilsen, \NP{\bf B147}(1979)277;\\
J. Harvey, P. Ramond and D. Reiss, \PL{\bf B92}{(1980)}{309};\\
I. Antoniadis, J.S. Hagelin, J. Ellis and D.V. Nanopoulos, \PL{\bf B205} (1988) 459;\\
G. K. Leontaris,  \PL{\bf B207}(1988  )447;\\
A. Faraggi,  \PR{\bf D47}(1993)5021;\\ 
Y. Nir and N. Seiberg, \PL{\bf B309}(1993)337;\\
L. Iba\~nez and G.G. Ross, \PL{\bf B332}(1994)100;\\
P. Binetruy and P. Ramond, \PL{\bf 350}(1995)49;\\
E. Papageorgiu, {\it Z.Phys.} {\bf C64}(1994) 509;\\
P. H. Frampton and O. C.W. Kong , \PRL {\bf 75}(1995)781 and {\bf 77}(1996)1699; \\
E. Dudas, C. Grojean, S. Pokorski and  C.A. Savoy, \NP{\bf B481}(1996)85;\\
B. Allanach, S. F. King, G. K. Leontaris and S. Lola, \PR{\bf D56}(1997)2632.
%%%%%%%%%%
\bibitem{kahler}
S. Ferrara, C. Kounnas, and F. Zwirner,
\NP{\bf B429}(1994)589 and references therein.
%%%%%%%%%%%
%\bibitem{kahlerexp} E. Witten, \PL{\bf B155}(1985)151.
%M. Cvetic, J. Louis and B. Ovrout, \PL {\bf B206}(1988)227;\\
%S. Ferrara and M.  Porrati, \PL{\bf  B216}(1989)289.
%%%%%%%%%%%%%%
\bibitem{trans} I. Antoniadis, E. Gava and K.S. Narain,
\PL{\bf B 283}(1992)209; \NP{\bf B383}(1992)93;\\ 
L. Ib\'a\~nez and D. L\"ust, \NP{\bf B283}(1992)305.
\bibitem{sav}E. Dudas, S. Pokorsky and C.A. Savoy,
\PL{\bf 369}(1996)255;\\
C.A. Savoy, FCNC inSUSY theories, Talk at HEP95 Euroconference, Brussels, July, 1995.
%%%%%%%%%%%%%
\bibitem{IR}L. Iba\~nez and G.G. Ross, as in \cite{u1}.
%%%%%%%%%%%%%
\bibitem{meg}
F. Gabianni and A. Masiero, \PL {\bf B 209}(1988)289;\\
 T. Kosmas, G.K. Leontaris and J.D. Vergados
\PL {\bf B 219}(1989) 457; {\it Prog.Part.Nucl.Phys.} {\bf 33}(1994)397;\\
R. Barbieri, L. Hall and A. Strumia, \NP{\bf B 445}(1995)219;\\
S. Dimopoulos and D. Sutter, \NP{\bf B}(1995).

\end{thebibliography}
\end{document}